\documentclass[aps,prl,amsmath,amssymb,showpacs,twocolumn,preprintnumbers,floatfix]{revtex4}

\usepackage{bm}
\usepackage{epsfig}
\usepackage{graphicx}
\usepackage{graphics}
\usepackage{amsmath}
\usepackage{amssymb}
\usepackage{amsfonts}
\usepackage{color}

\newcommand{\be}{\begin{equation}}
\newcommand{\ee}{\end{equation}}
\newcommand{\ba}{\begin{eqnarray}}
\newcommand{\ea}{\end{eqnarray}}
\newcommand{\br}{\begin{array}}
\newcommand{\er}{\end{array}}

\begin{document}

\author{Xiaoxu~Guan}\email{xiaoxu.guan@drake.edu}
\author{Klaus~Bartschat}\email{klaus.bartschat@drake.edu}
\affiliation{Department of Physics and Astronomy, Drake University, Des Moines, Iowa 50311, USA}
\date{\today}

\title{\bf Complete break-up of the helium atom by proton and antiproton impact}

\begin{abstract}
We present a fully {\it ab initio}, non-perturbative, time-dependent approach to 
describe single and double ionization of helium by proton and antiproton impact.
A flexible and accurate finite-element discrete-variable-representation is applied 
to discretize the problem on the radial grid in spherical coordinates. Good 
agreement with the most recent experimental data for absolute angle-integrated 
cross sections is obtained over a wide range of incident projectile energies 
between $3$~keV and $6$~MeV. Furthermore, angle-differential cross sections
for two-electron ejection are predicted for a proton impact energy of $6$~MeV. 
Finally, the time evaluation of the ionization process is portrayed by displaying
the electron density as a function of the projectile location.
\end{abstract}

\maketitle
Ionization of helium by slow antiproton impact has received considerable attention in recent 
months, from experimentalists~\cite{Knudsen2008,Knudsen2009} and theorists alike~\cite{Foster2008,Foster2008-2}.
As a fundamental, strongly correlated few-body
collision system, it is at the heart of examining
the reliability of the most advanced computational methods for atomic collision
processes.  Using antiprotons
has the major advantage of eliminating the complicated charge-exchange process that is competing 
with single ionization in the case of proton impact.

Of particular interest regarding the validity of theoretical approaches is the
low-energy region,
where the projectile speed~$|\bm{v}|$ is sufficiently slow that the
Massey perturbation
parameter~$|Z_p|/|\bm{v}|$ \cite{McGuire1997},
where $Z_p$ is the charge of the projectile, is becoming larger than unity. Note that
$|Z_p|/|\bm{v}| = 1$ for 
incident (anti)protons with energy $E_i \simeq 25\,$keV.  Consequently,
perturbative treatments based on 
one or even a few terms of the Born series are entirely
inappropriate for the problem at this and lower energies.  Instead, non-perturbative treatments, either based upon the close-coupling expansion
with an appropriate set of basis functions~\cite{Kirchner2002,Keim2003}
or the direct 
solution of the time-dependent Schr\"odinger equation (TDSE)
on a numerical space-time grid~\cite{Schultz2003,Foster2008,Foster2008-2} are required. The latter method is often
referred to as the time-dependent close-coupling (TDCC) approach.

The interaction between the charged projectile and the target can also be
viewed as a temporal ultrafast half-cycle-like pulse, which is responsible for 
the (multiple) ionization process of the target~\cite{Ullrich2003}. The full width at half maximum
of such a pulse can be estimated as $2\sqrt{3}\,b/|{\bm v}_p|$. For
(anti)proton impact, this is 
of the order $10$~atto\-seconds for ``intermediate'' energies around $100\,$keV. Further\-more,
the peak strength of the electric field ${\cal U}_{\max}$ is
of the \hbox{order~$Z_p/|\bm{b}|$}. Even for (anti)protons as the lightest ``heavy'' particles, ${\cal U}_{\max}$ is
approximately two atomic units at an impact parameter $|\bm{b}|\simeq 1\,a_0$, where $a_0 = 0.529 \times 10^{-10}\,$m is 
the Bohr radius. This results in peak intensities of the half-cycle pulse as high as
$10^{17}\,$W/cm$^2$.  Consequently, charged-ion impact presents 
an interesting alternative to intense laser-pulse techniques in the study of 
ionization processes in strong electromagnetic fields within an
ultra\-short time window~\cite{Ullrich2003}.

Various experimental datasets for single and double ionization of
helium by antiproton impact were published over the years. While there is generally
good agreement at high incident energies,
both between different experimental datasets and predictions from various theoretical models, the situation
is much less clear for incident energies of 20$\,$keV and below.  Most recently, Knudsen {\it et al.}~\cite{Knudsen2008}
published results for single ionization of helium by antiproton impact that
differed substantially from those
obtained earlier by Andersen {\it et al.}~\cite{Andersen1990} and Hvelplund {\it et al.}~\cite{Hvelplund1994}.
While the recent experimental data of Knudsen {\it et al.}~\cite{Knudsen2008} were reproduced fairly well by
various non-perturbative calculations~\cite{Kirchner2002, Keim2003,
Foster2008}, discrepancies
of $15\%$ or more still remained.

Here we report results from a fully {\em ab initio} numerical study of the helium break-up 
problem by antiproton and proton impact over a wide range of incident energies
between
$3\,$keV to $6\,$MeV, as well as the corresponding total and double-differential
cross sections (DDCS). 
Our approach, whose validity is not restricted to a particular projectile energy range, thus provides a unique opportunity 
for studying the multiple ionization dynamics induced by charged ions from weak to strong
perturbations. The configuration space of the target electrons is
discretized via a finite-element discrete-variable 
representation (FE-DVR).
This highly flexible and accurate grid-based approach combines the
numerical advantages of basis-function expansions in small intervals with an
easily adaptable spatial grid to
account for the radial dependence of the electronic wave\-function close to and
far away from the nucleus.  

The dynamical response of the system is obtained by propagating the initial wave\-packet, defined on
the DVR grid\-points, through a recently developed time-dependent Arnoldi-Lanczos
algorithm~\cite{Guan2008,Guan2009-CPC}.
The collision system is governed by 
the time-dependent Hamiltonian
\begin{equation}
{\cal H}(t)=-\frac{\bm{\nabla}^2_1}{2}-\frac{\bm{\nabla}^2_2}{2}
-\frac{Z_t}{r_1}-\frac{Z_t}{r_2}+\frac{1}{|\bm{r}_1-\bm{r}_2|}+{\cal U}_p(t).
\end{equation}
The essential complexity, compared to previous work
involving a spatially homogeneous
laser field~\cite{Guan2008}, is the representation of the
two-body Coulomb interaction 
\begin{eqnarray} 
{\cal U}_p(t) &=& - Z_p/|\bm{r}_1-\bm{R}(t)| - Z_p/|\bm{r}_2-\bm{R}(t)| \nonumber \\ 
       &&\hspace{-1.5cm}=-Z_p\sum_{\lambda_1
q_1}\frac{4\pi}{2\lambda_1+1}
                  \frac{[r_1,R(t)]_{<}^{\lambda_1}}{[r_1,R(t)]_{>}^{\lambda_1+1}} 
                  Y_{\lambda_1 q_1}^{*}(\hat{\bm{r}}_1)Y_{\lambda_1 q_1} (\hat{\bm{R}}(t)) \nonumber \\ 
              &&\hspace{-1.5cm}-Z_p\sum_{\lambda_2
q_2}\frac{4\pi}{2\lambda_2+1}
\frac{[r_2,R(t)]_{<}^{\lambda_2}}{[r_2,R(t)]_{>}^{\lambda_2+1}}
                  Y_{\lambda_2 q_2}^{*}(\hat{\bm{r}}_2)Y_{\lambda_2
q_2}(\hat{\bm{R}}(t))
\end{eqnarray}
between the projectile and the target.  
Here $\bm{r}_1$ and $\bm{r}_2$ are the coordinates of the two helium electrons,
$\bm{R}(t)$ is the coordinate of the projectile, and $[x,y]_{<(>)} \equiv \min(\max)\{x,y\}$. All these coordinates are
defined relative to the He$^{2+}$ ion, which is fixed at the origin.

We use a straight-line
trajectory \hbox{$\bm{R}(t)=\bm{b}+(\bm{d}_0-\bm{v}t)$} for the incident
projectile starting at $\bm{d}_0$ at an impact
parameter~$\bm{b}$.  
This is the only physical approximation made in our treatment. It should be sufficiently accurate
even at a projectile energy as low as $3$~keV, which is the lowest energy
considered in this work. At this energy and \hbox{$|\bm{b}|\simeq 1\,a_0$}, which is 
the impact-parameter regime with the largest contribution to the cross sections, the
scattering angle in the laboratory system is estimated to be merely $1^\circ$. This results in
a small relative momentum transfer $|\bm{q}|/|\bm{P}_i|$ of $1.8\%$, where $\bm{P}_i$ is the
projectile's initial momentum.

As a result of using a classical trajectory for the projectile, 
the total angular momentum $L$, magnetic quantum number $M$, and parity $\Pi$
of the collision system are no longer conserved quantities that could be taken advantage of in a fully quantal
partial-wave expansion. However, there is still one conserved quantity, namely the reflection symmetry of the 
electronic wave\-function with respect to the collision plane.  The latter is defined by the trajectory of
the incident projectile and the impact parameter~$\mbox{\boldmath$b$}$.

We explicitly build this reflection symmetry into our FE-DVR wave\-function by
writing 
\vspace{-3mm}
\begin{widetext}
\vspace{-5mm}
  \begin{eqnarray}
  \Psi(\bm{r}_1,\bm{r}_2,t)  
      & = & 
          \sum_{LM,l_1l_2}\sum_{j < i}
          \bigg[f_i(r_1)f_j(r_2)C_{l_1l_2LM}^{ij}(t)+
        (-1)^{l_1+l_2-L}f_j(r_1)f_i(r_2)C_{l_2l_1LM}^{ij}(t)\bigg]  
          {\cal G}_{l_1l_2}^{LM}(\hat{\bm{r}}_1,\hat{\bm{r}}_2)\nonumber
\\
      & + &
          \sum_{LM,l_1\leqslant l_2}\sum_{i}
          f_i(r_1)f_i(r_2)
C_{l_1l_2LM}^{ii}(t)\frac{1}{1+\delta_{l_1l_2}} 
          \bigg({\cal G}_{l_1l_2}^{LM}(\hat{\bm{r}}_1,\hat{\bm{r}}_2)+
          (-1)^{l_1+l_2-L}
          {\cal G}_{l_2l_1}^{LM}(\hat{\bm{r}}_1,\hat{\bm{r}}_2)\bigg).
          \label{eq:globalwv}
  \end{eqnarray}
  \vspace{-1mm}
\end{widetext}
\vspace{-3mm}
Instead of ordinary coupled spherical harmonics ${\cal
Y}_{l_1l_2}^{LM}(\hat{\bm{r}}_1,\hat{\bm{r}}_2)$, 
we introduced the  angular basis 
\begin{equation}
\mathcal{G}_{l_1l_2}^{LM}(\hat{\bm{r}}_1,\hat{\bm{r}}_2) \equiv
\sqrt{\frac{2}{1+\delta_{M0}}}\mbox{Re}\Big[{\cal
Y}_{l_1l_2}^{LM}(\hat{\bm{r}}_1,\hat{\bm{r}}_2) \Big].
\label{newang}
\end{equation}
These basis functions are normalized according to 
$\langle{\cal G}_{l_1l_2}^{LM}| {\cal
G}_{l'_1l'_2}^{L'M'}\rangle=\delta_{l_1l'_1}\delta_{l_2l'_2}
\delta_ { LL' } \delta_
{ MM'} $.
Using the reflection symmetry relates the wave\-functions for the magnetic quantum numbers~$M$ and $-M$,
in the same way as it does for electron-impact excitation and co\-planar ionization and ionization-excitation
processes~\cite{PAO}.
Consequently, we only need to include $M \ge 0$ in Eq.~(\ref{eq:globalwv}) and thus the size of the
problem is reduced to nearly half of what it would be without 
adopting the above symmetry. Taking advantage of the reflection symmetry, 
we use $55$ angular partial
waves generated by setting $(l_{1},l_{2},L,M)_{\max}=(3,3,3,3)$ to expand the
wave function in Eq.~(\ref{eq:globalwv}). This is equivalent to the $101$ partial
waves used in Ref.~\cite{Foster2008}.

To calculate the angle- and energy-integrated cross sections, $399$ DVR grid points were set 
up in a spatial box of \hbox{$r_{\max}=80\,a_0$},  
while a smaller step size and $799$ points were employed for the DDCS.
We also extended the box size to \hbox{$160\,a_0$} in this case to
ensure converged results.
 
\begin{figure}[bht]
\centering
\epsfig{file=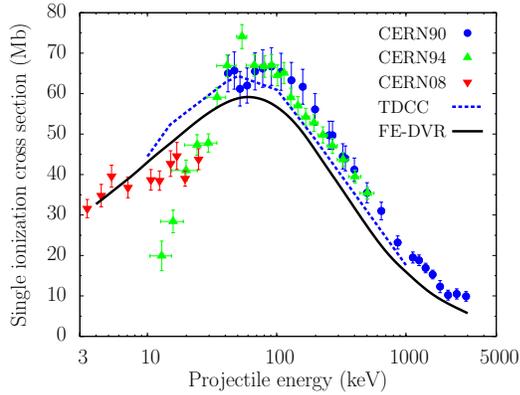,width=6.9cm,clip=}
\caption{(Color online) Cross section for single ionization of helium
by antiproton impact.  
Experimental data
obtained at CERN by Andersen {\it et al.}~\cite{Andersen1990} (CERN90),
Hvelplund {\it et al.}~\cite{Hvelplund1994} (CERN94),
and Knudsen {\it et al.}~\cite{Knudsen2008} (CERN08) are compared with
TDCC~\cite{Foster2008}
and the present FE-DVR predictions.}
\label{fig:sion-pbar}
\end{figure}

The total cross section (angle- and energy-integrated) is obtained as 
\begin{equation}
\sigma(E_i) = 2\pi\int_{0}^{+\infty} P(b,E_i)bdb, 
\end{equation}
where $P(b,E_i)=\int |\langle\Phi_{k_1k_2}| \Psi(t)\rangle|^2dk_1dk_2$
represents the probability for single or double ionization at fixed values of $|\bm{b}|$ and $E_i$.
Figure~\ref{fig:sion-pbar} exhibits our results for single
ionization of helium
by antiproton impact for
projectile energies between $3\,$keV and $5\,$MeV.  In the theoretically most difficult low-energy regime,
we obtain excellent agreement with the most recent experimental data of Knudsen {\it et al.}~\cite{Knudsen2008}.  For projectile
energies of $20\,$keV and above, our results are slightly lower than those of Foster {\it et al.}~\cite{Foster2008} and thus
in better agreement with experiment. On the other hand, most experimental data near the cross section maximum
around 100$\,$keV and beyond lie above both our results and those of Foster {\it et al}.  Having performed
extensive convergence checks, we are confident in the numerical accuracy of our predictions and currently have no explanation for
the remaining discrepancies.

\begin{figure}[tbh]
\centering
\epsfig{file=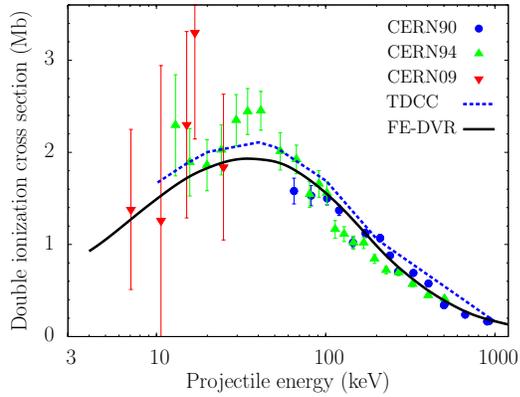,width=6.9cm,clip=}
\caption{(Color online) Cross section for double ionization of helium
by antiproton impact. The experimental data
obtained at CERN by Andersen {\it et al.}~\cite{Andersen1990} (CERN90), Hvelplund {\it et al.}~\cite{Hvelplund1994} (CERN94),
and Knudsen {\it et al.}~\cite{Knudsen2009} (CERN09) are compared with TDCC~\cite{Foster2008}
and the present FE-DVR predictions.}
\label{fig:dion-pbar}
\end{figure}

Figure~\ref{fig:dion-pbar} depicts the corresponding results for 
the double ionization process.   
Although the size of the experimental error bars \cite{Knudsen2009} is
substantial and thus limits the meaning of comparing
the absolute numbers between theory and experiment,
we note that our results are in excellent agreement with those of Foster {\it et al.}~\cite{Foster2008} in the limited range
of projectile energies where their data are available.

\begin{figure}[tbh]
\epsfig{file=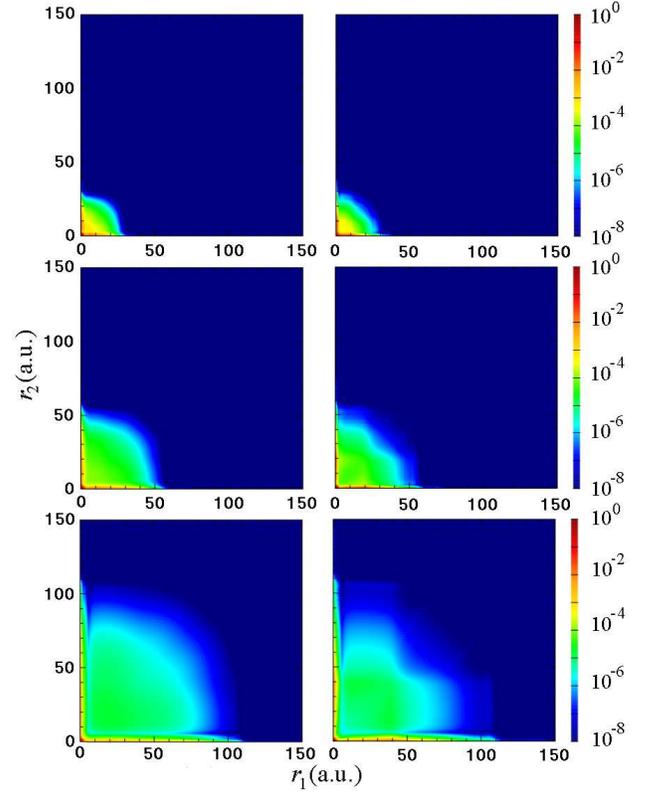,width=8.2cm,clip=}\\
\caption{(Color online) Radial electron density after 
antiproton (left panels) and proton (right panels) impact on helium at an
energy of
$100\,$keV for an impact parameter of $0.5\,a_0$. Starting from an initial distance of $-50\, a_0$,
the positions of the projectile shown in the snapshots, from top to bottom,
are $+10$, $+20$, and $+40\,a_0$ relative to the
center of the target, respectively.}
\label{fig:radial1}
\end{figure}

To gain further insight into the dependence of the joint two-electron
response on the sign of the projectile charge, we show in the left panel of Fig.~\ref{fig:radial1} an example
for $100\,$keV antiproton impact on helium
at an impact parameter of $0.5\,a_0$, while the right panel shows the
corresponding results for 
proton impact.  We first note that not much happens until the
projectile is already 10$\,a_0$ passed the target (top panels). For this case,
and also a little
later when the projectile is
$20\,a_0$ behind the target (center panels), the radial electron densities show a substantial double-ionization
component, whose signature is a significant density at large values of both $r_1$ and $r_2$. The most energetic electrons have
moved to about $25$ and $50\,a_0$, respectively.  Finally, when the projectile is $40\,a_0$
beyond the target (bottom panels), we see the characteristics of both double
and single (characterized by significant densities when only $r_1$ or $r_2$ is large) ionization processes developing. 
By this stage, the most energetic electrons
have moved as far away as $100\,a_0$ from the He$^{2+}$ center.
When comparing the electron densities for the two projectile charges, we recall
that the antiproton cannot
capture any of the electrons and in fact pushes them away, whereas the capture
channel may be important for proton impact. 
While the single-ionization signals for the two projectiles resemble each other, the major difference between 
the results for the two projectiles occurs in the double-ionization region of $r_1 \approx r_2$.
Apparently the proton is trying to attract at least one electron and hence
causes a reduction in the probability for both electrons moving out with the
same speed.

\begin{figure}[tbh]
\epsfig{file=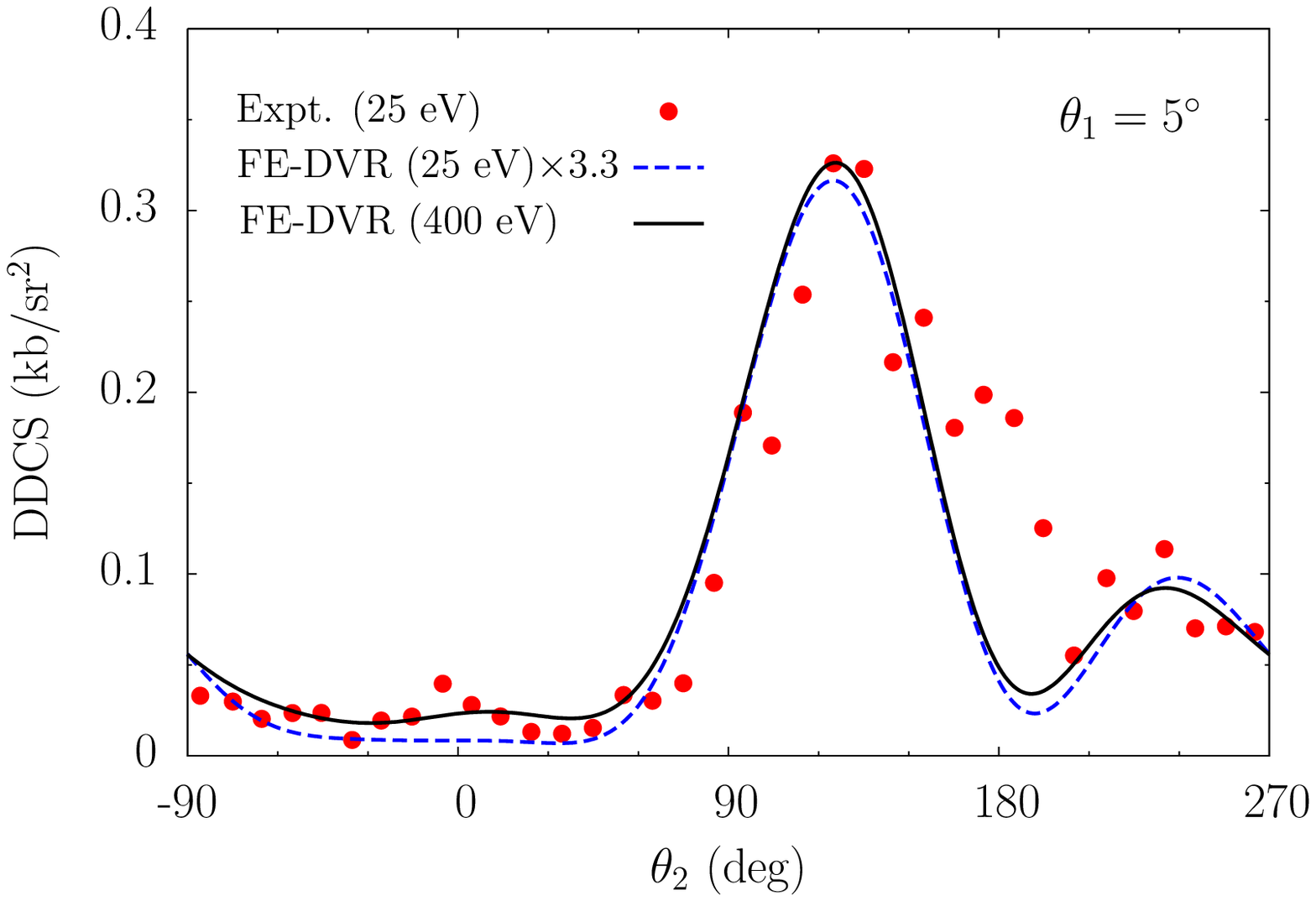,width=6.0cm,clip=}\\
\epsfig{file=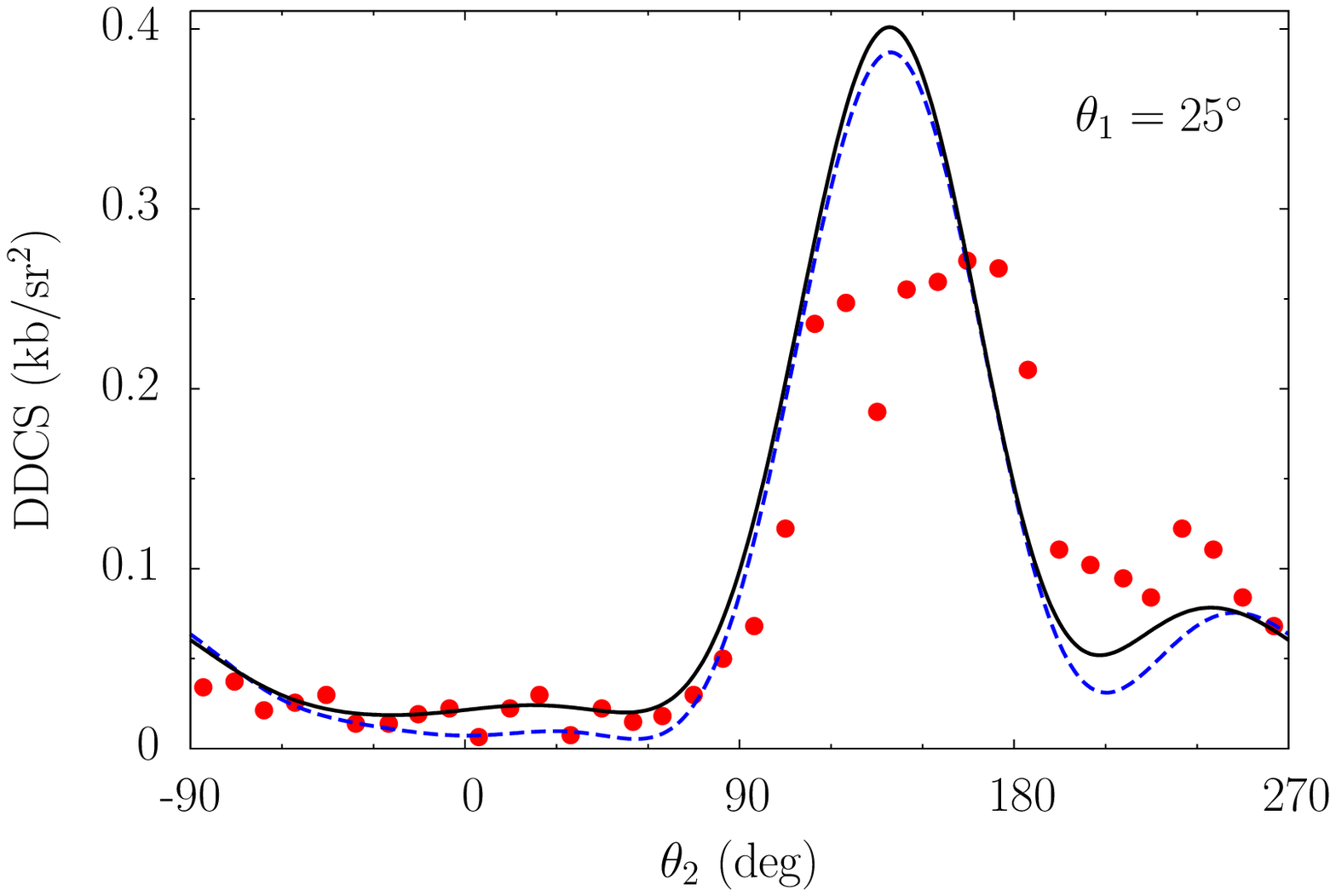,width=6.0cm,clip=}\\
\epsfig{file=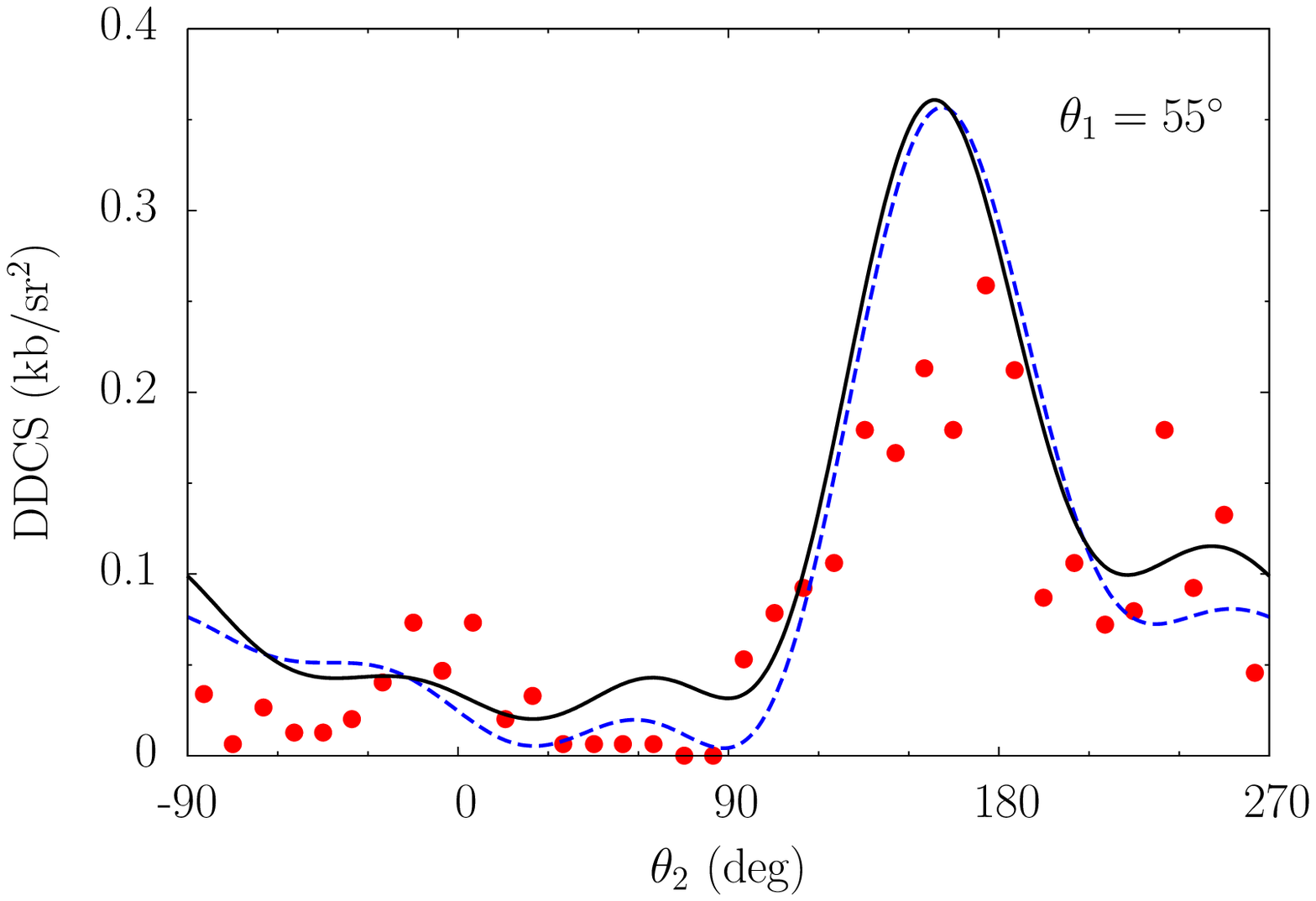,width=6.0cm,clip=}\\
\caption{(Color online) DDCS for proton impact 
double ionization of helium for an 
incident energy of $6\,$MeV, a fixed detection angle~$\theta_1$ for one of the electrons, and a
variable detection angle~$\theta_2$ for the second electron.  
The experimental measurements of Refs.~\cite{Fischer2003,Schulz2005} were
normalized to the converged 
DDCS at the large peak for $\theta_1=5^\circ$.}   
\label{fig:ddcs}
\end{figure}

In order to portray the two-electron emission in a more quantitative way,
it is important to consider angle-resolved cross sections, e.g., the DDCS for 
two-electron ejection without observing the electrons' individual energies.  This 
particular DDCS is obtained as 
\begin{eqnarray}
&&\hspace{-1cm}\frac{d^2\sigma}{d\hat{\bm{k}}_1d\hat{\bm{k}}_2}
=2\pi\int_{0}^{+\infty} \!\!\! bdb \int_{0}^{+\infty} \!\!\! dk_1 \int_{0}^{+\infty} \!\!\! dk_2 
\bigg|\sum_{LM,l_1l_2}(-i)^{l_1+l_2} \nonumber \\ 
&&\hspace{5mm}
\times \, e^{i(\sigma_{l_1}+\sigma_{l_2})}
{\cal F}_{b,l_1l_2}^{LM}(k_1,k_2){\cal
G}_{l_1l_2}^{LM}(\hat{\bm{k}}_1,\hat{\bm{k}}_2)
\bigg|^2,
\label{eq:ddcs}
\end{eqnarray}
where $\sigma_{l}$ denotes the Coulomb phase and ${\cal
F}_{b,l_1l_2}^{LM}(k_1,k_2)$ is the partial-wave amplitude in momentum space.

Figure~\ref{fig:ddcs} exhibits such DDCS results for proton impact
double ionization of helium for an incident energy of $6\,$MeV.
The results are for the coplanar geometry, where the momentum transfer and the
momentum vectors of the two ejected electrons all lie in the same plane.
We compare our FE-DVR predictions for cuts with fixed values of $\theta_1$ 
that were generated by Foster {\it et al.}~\cite{Foster2008-2}
from the original experimental data of Fischer {\it et al}.~\cite{Fischer2003}
after the analysis by Schulz {\it et al}.~\cite{Schulz2005}. 

Note that only ejected electrons with energies up to 25$\,$eV were observed 
experimentally, while converged results for the DDCS defined in Eq.~(\ref{eq:ddcs}) 
require energies up to $\simeq 400\,$eV to be counted in this case. 
Although the measured signal thus only corresponds 
to about one third of the converged DDCS, we see that the shape of the DDCS is 
essentially determined even with the cutoff at 25$\,$eV.

We see good agreement with the experimental data and clearly reproduce the systematic shift 
of the large peak to the right with increasing values of~$\theta_1$.  Furthermore, the angle 
between $\theta_1$ and the maximum in the DDCS as a function of~$\theta_2$ decreases slowly 
from about~120$^\circ$ for \hbox{$\theta_1 = 5^\circ$} to about~100$^\circ$ 
for \hbox{$\theta_1 = 55^\circ$}.  
Similar results were published by Foster {\it et al.}~\cite{Foster2008-2}, but
both the graphical presentation and the theoretical magnitudes given in
Fig.~2 of their paper contain errors~\cite{Colgan2009}. 

Finally, we obtained values of $3.54$ and $8.97\times 10^{-3}$ Mb, respectively,
for the total cross sections for single and double ionization.
Not surprisingly for such a high projectile energy, 
their ratio of $\simeq 400$ is consistent with what one would expect from the
shake-off mechanism. 

In summary, we have investigated the complete break-up problem of helium by
proton and antiproton impact by using a time-dependent non-perturbative FE-DVR 
approach. This is a prime example of a highly
correlated four-body Coulomb process, whose description remains a major
theoretical and computational challenge. Generally good agreement with the
latest sets of experimental data for both integrated and differential cross
sections was obtained. At lower projectile energies ($<20\,$keV), our 
antiproton results clearly show that the cross section for double ionization 
decreases with decreasing projectile energy. For the angle-resolved DDCS
presented here, converged results require to account for contributions from
ejected electrons with energies of several hundred~eV, whereas the angular 
dependence is essentially established by low-energy electrons with energies of
less than $\simeq 25\,$eV.

\vspace{2.0truemm}
We thank Prof.~Helge Knudsen for sending us experimental data in numerical form and
helpful discussions and Dr.~James Colgan for clarifying problems with
Ref.~\cite{Foster2008-2}. This work was supported by the NSF under grant 
PHY-0757755 and supercomputer resources through the Teragrid allocation TG-PHY090031.

\end{document}